\documentclass[preprint,showpacs,preprintnumbers,amsmath,amssymb]{revtex4}
\usepackage{graphicx}
\begin{document}


\title{The Influence of Electro-Mechanical Effects on Resonant Electron
Tunneling Through Small Carbon Nano-Peapods}


\author{I. V. Krive$^{1,2}$, R. Ferone$^1$, R.~I.~Shekhter$^1$, M. Jonson$^{1,3}$, P. Utko$^4$, J. Nyg{\aa}rd$^4$ }
\affiliation{$^1$Department of Physics, University of Gothenburg, SE-412
96 G\"{o}teborg, Sweden\\
$^2$B.~I.~Verkin Institute for Low
Temperature Physics and Engineering, 47 Lenin Avenue, 61103 Kharkov,
Ukraine\\
$^3$School of Engineering and Physical Sciences, Heriot-Watt University,
Edinburgh EH14 4AS, Scotland, UK\\
$^4$Niels Bohr Institute \& Nano-Science Center,
University of Copenhagen, DK-2100 Copenhagen, Denmark}

\begin{abstract}
The influence of a fullerene molecule trapped inside a single-wall carbon
nanotube on resonant electron transport at low temperatures and strong polaronic coupling
is theoretically discussed. Strong peak to peak fluctuations and anomalous
temperature behavior of conductance amplitudes are predicted and investigated. The influence of the
chiral properties of carbon nanotubes on transport is also studied.
\end{abstract}

\pacs{73.63.Fg, 73.23.-b, 73.63.-b}

 \maketitle

\section{Introduction}
Carbon nano-peapods were discovered almost one decade ago \cite{1}
(see also the review papers Refs.~\cite{2,Krive,Month}). They have been
proven to be novel graphitic structures with electrical and
mechanical properties that can not be reduced to the sum of
properties of their subcomponents (an empty carbon nanotube and
individual fullerene molecules). Peapods, a composite system
consisting of a rigid single-wall nanotube (SWNT) cage and an inner
mechanically soft chain of fullerenes, can exhibit interesting
nano-electro-mechanical properties if the coupling of delocalized
electron states on the tube with the fullerene molecular orbitals is
sufficiently strong. Buckyballs inside a SWNT are neutral objects
and one can not expect any direct electrical influence of
encapsulated molecules on electron transport along a nanotube shell.
However, STM spectroscopy of semiconducting peapods \cite{3}
showed that the local
electron density of states in carbon nanotubes is strongly modified
by encapsulated $C_{60}$ molecules. These measurements can be
theoretically interpreted in terms of a strong hybridization of
electronic states on the nanotube surface with the lowest unoccupied
molecular orbitals (LUMO) of the encaged $C_{60}$. The corresponding
hybridization strength was estimated in Refs.~\cite{3,4} to be close
to $1$ eV. This energy is comparable to other characteristic energies
of SWNTs and this means that virtual hopping of delocalized
electrons from the SWNT to the local LUMO states is possible and can
strongly influence the electronic properties of the SWNTs.
Even though all samples
in the cited experiment were semiconducting nanotubes,
one could expect from
theoretical considerations an analogous strong
hybridization at least for {\em chiral} metallic nanotubes (see
Ref.~\cite{4} where a theory of scanning tunneling spectroscopy of
long  $C_{60}$ peapods was developed). Notice that such a strong hybridization
between fullerene-derived and nanotube-derived energy levels in peapods
could shift the LUMO states even below the Fermi energy of the SWNT. This
possibility was however ruled out in recent photoemission experiments, \cite{Shiozawa}.

In this work we suggest that a strong coupling between the electronic states
on the SWNT shell and the localized states on the fullerenes affects
resonant tunneling transport of electrons through the metallic peapod system
in a manner that is experimentally observable. In particular, we
propose that  resonant electron transport due to polaronic effects
provides a method for
studying the fullerene nanomechanical dynamics through the
temperature dependence of the resonant conductance peaks. This
should be readily observable since the characteristic temperature
scale for the strong polaronic modification of the peaks is
set by the fullerene vibration frequencies and may be very low
compared to typical electronic energies.

Recent experiments with high-quality peapods
demonstrated Coulomb blockade and Kondo physics at low temperatures, \cite{USA,Utko}.
The spacings of conductance peaks was found to be regular over a
wide range of gate voltages, indicating that the Constant Interaction
Model of Coulomb blockade (see e.g. Ref.~\cite{CIM}) applies for these dots. The peak amplitudes, on the other hand, strongly
 fluctuate from peak to peak. In the case of an empty metallic
SWNT both peak spacings and amplitudes should ideally vary
slowly with the gate voltage, except for variations of level
spacings dictated by the particular SWNT bandstructure which leads
to four-electron shells. While earlier experiments on empty
nanotubes have exhibited strongly fluctuating peak conductances
(see e.g., Ref.~\cite{Jesper}), defect-free SWNTs have indeed shown
remarkably gate-independent peak amplitudes \cite{Cao,Sapmaz}.

We show here that for carbon nano-peapods, electromechanical effects
can lead to strong fluctuations of the Coulomb peaks. The mechanism
will be effective even with defect-free nanotube shells. This phenomenon
may explain the strongly fluctuating peak amplitudes seen in
peapod experiments \cite{USA,Utko} - detailed studies of Coulomb blockade peak
fluctuations could thus reveal polaronic effects in quantum dots. Molecular conductors
constitute another important class for which the vibronic coupling
is also nearly always present, and for which it plays a fundamental role in the
study of the transport properties, see \cite{joachim, galperin} and references therein.

The paper is organised as follows. In the following Section \ref{spectrum}, we investigate how the presence
of even one encapsulated fullerene molecule, treated as a short-range scattering potential,
can influence the electron energy spectrum of the SWNT by a strong polaronic effect. This will allow us to define the Hamiltonian of the peapod in an
independent electron model. In Section \ref{transport},
the current through the peapod and hence its conductance are evaluated and studied, in
this regime of strong polaronic coupling. Since the fullerene scattering
potential in reality has an intermediate range, we
discuss in Section \ref{Graphene} how the peapod conductance is affected in the
opposite limit of a long-range fullerene-induced scattering potential. Here, the appropriate
model for the SWNT electrons involves the use of the Dirac equation.
Finally, in Section V we present our conclusions.

%

\section{Energy spectrum in presence of a short-range scattering potential: the model}\label{spectrum}

The position of an individual buckyball inside the tube, for a partially filled peapod,
is not fixed, and mechanical motion of the
fullerene is excited in the process of virtual electron transitions
onto and off the fullerene LUMO level. The effective excitation energy
$\hbar\omega_0$ strongly depends on the buckyball confinement
potential. For vibrations in the nearly harmonic (for small
displacements) transverse Girifalco potential \cite{5,6} the corresponding
energy $\hbar\omega_\mathrm t$ is of the order of meV's.
Notice that electro-mechanical vibrations of $C_{60}$ in the van der
Waals-like potential between two gold electrodes have already been
measured ($\hbar\omega\simeq 5$ meV) in tunneling experiments
\cite{7}. The longitudinal motion of $C_{60}$ inside the tube is less
restricted and the corresponding excitation energy $\hbar\omega_\mathrm l$
could be much smaller. It is worth to notice here that the nanotube shell of
peapods can in fact be quite full of large ($\sim 1$ nm) holes in the tube
walls. Only a small
number of openings, used before to fill the tube, are healed in the post processing, \cite{thanks}. The
average spacing between the holes can be estimated to be $\sim 5-10$ nm,  \cite{2}. The holes
produce a repulsive potential for fullerenes inside the tube, which we
will model in our analytical calculations as a harmonic confinement potential
with a characteristic energy $\hbar\omega_0\sim 0.1$ meV,
corresponding roughly to the zero-point energy of a $C_{60}$ confined
to a length equal to the estimated average spacing between holes.

We start with a single encapsulated $C_{60}$ placed at a distance
$l$ from the nanotube end. The virtual processes of electron hopping
onto and off the fullerene LUMO level (shifted upwards by the
charging energy $\sim e^2/2d$, where $d$ is the fullerene diameter)
produce a local attractive potential for conduction electrons on the
tube, \cite{4}. The first question one has to consider in order to
formulate a theoretical model for the problem is whether this
potential is short- or long-range with respect to its effect on
electrons in metallic nanotubes. If the characteristic length of the
scattering potential due to the fullerene is defined by its diameter,
of the order of $1$ nm, its range is determined by a comparison with the
Fermi wavelength, $\lambda_\mathrm F$, of electrons in a metallic SWNT. Since
$\lambda_\mathrm{F}\sim 7$ \AA, \cite{lambdaF}, and hence is remarkable close to the fullerene diameter,
the range of the potential is of intermediate range. It is therefore not
a priori clear which of the two different limits, where an analytical
investigation is possible, is the more appropriate one and we shall
therefore discuss both.

In the first limit, the scattering potential is described
as a short-range potential, which means that the hybridization-induced
scattering potential is able to strongly backscatter electrons with a large
momentum transfer, $\delta k \simeq 2k_\mathrm F$. This is the regime studied in this
Section by means of a single-electron model. In the opposite limit the
scattering potential is described as a long-range potential when the real band
structure of the SWNT becomes important. This limit is hence discussed
in Section \ref{Graphene} using the Dirac equation.


A short-range hybridization-induced scattering potential with the
hybridization strength $t_\mathrm h\simeq 1$ eV changes the distribution
of level spacings in the conduction band of a finite-sized
nanotube. By solving the Schr\"{o}dinger equation on a finite interval
($L$) with a $\delta$-function scatterer placed at point $x=l$ one
gets the spectrum equation
\begin{equation}  \label {ss}
k\sin kL = U(k)\sin kl\sin k(L-l) ,
\end{equation}
where $k$ is the electron wave vector,
($\varepsilon=\hbar^2k^2/2m$), and the amplitude of the local scattering
potential in our case represents the hybridization potential of
Ref.~\cite{4}. In our notation
\begin{equation} \label{Uk}
U(k)=\left(\frac{k_\mathrm F^2}{\varepsilon_\mathrm F}\right)
V_\mathrm h(\varepsilon)\;\;,\;\;V_\mathrm h(\varepsilon)=
-\frac{|t_\mathrm h|^2d}{E_0-\varepsilon} \;.
\end{equation}
Here, $d\simeq\lambda_\mathrm F$ and $E_0>\varepsilon$ is the energy of the
hybridized side-level ($E_0\simeq 2\varepsilon_\mathrm F$ if one takes
into account the charging energy for electron hopping to the $C_{60}$
LUMO level).

The energy spectrum of Eqs. (\ref{ss}) and (\ref{Uk}) when $l/L$ is
not a small rational number is irregular for energies not far from
the Fermi energy (see FIG. 1),  and indicates certain deviations
from the mean level spacing $\Delta_L=\pi\hbar v_\mathrm F/L$
(notice that due to the two identical conduction bands in a SWNT its
mean level spacing is $\Delta_L/2$). However, as one can see from
FIG. 1, for a single scatterer the "randomization" is not pronounced
even for energies close to the Fermi energy. About $40\%$ of the
levels deviate from the mean level spacing less than $18\%$, and
among these levels $75\%$ deviate less than $12\%$.

\begin{figure}
\centering
\includegraphics[scale=0.9]{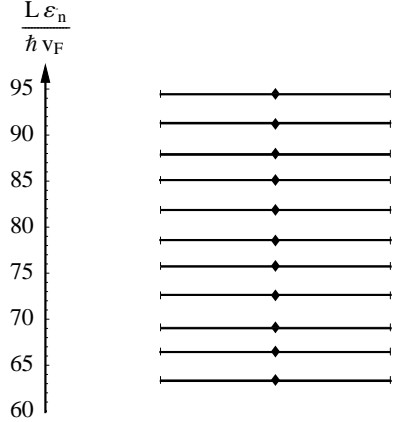}
\caption{Energy levels, $\varepsilon_n L/\hbar v_\mathrm F$, for a
peapod with just one fullerene inside. Unlike an empty SWNT, the
spectrum is not uniform; still, for a single scatterer (fullerene),
the randomization is not pronounced. The Fermi energy
$\varepsilon_F$ is such that $\varepsilon_F L/\hbar v_\mathrm F\sim
90$.}
\end{figure}

A shift $\delta x$ in the position of a scatterer leads to
shifts of all quantized energy levels
\begin{equation} \label{delta}
\delta\varepsilon_n=\hbar v_\mathrm{F}\frac{\partial k_n}{\partial x}
\delta x ,
\end{equation}
where the derivative of electron momentum with respect to
scatterer position can easily be found from Eq. (\ref{ss}).

In order to get analytical results we will model the potential that
confines the longitudinal motion of our scatterer (buckyball inside
a tube) as a harmonic potential. In the adiabatic limit,
$\omega_0\ll v_\mathrm F/L$, when the electron spectrum is well
defined for any position of the vibrating impurity, the total
Hamiltonian for a two-terminal conduction geometry reads$^1$
\footnotetext[1]{The total Hamiltonian ${\cal H}$ in Eq.
\ref{hamilt} does not take into account the scattering potential
represented by the holes on the tube shell. The possible
consequences of such a issue are discussed in the conclusions}:
\begin{equation}\label{hamilt}
{\cal H}= H_\mathrm{L}+H_\mathrm{R}+H_{\mathrm T,\mathrm{L}}+H_{\mathrm T,\mathrm{R}}+H_\mathrm{QD}\,
\end{equation}
where $H_\alpha=\sum_{k_\alpha}\varepsilon_{k_\alpha} c^\dagger_{k_\alpha} c_{k_\alpha}$
is the Hamiltonian for the leads and $\alpha={\mathrm{L,R}}$ for left and right reservoir respectively.
The quantum dot (QD) Hamiltonian reads:
\begin{equation} \label{Ham}
H_\mathrm{QD}=\sum_n\varepsilon_nc_n^{\dagger}c_n
+\sum_nV_n(L,l)c_n^{\dagger}c_n(b^{\dagger}+b)
+\hbar\omega_0b^{\dagger}b ,
\end{equation}
where $\varepsilon_n$ is the set of levels shown in FIG. 1,
$\hbar\omega_0$ is the vibration energy quantum, $V_n(L,l)=\hbar
v_Fx_0(\partial k_n/\partial l)$ and $x_0$ is the amplitude of
zero-point fluctuations of the bosonic field ($x_0\simeq 0.2$ \AA\,  for
$C_{60}$ and for an energy value $\hbar\omega_0\sim 0.1\; \mathrm{meV}$); $c_n (c_n^{\dagger})$ and $b (b^{\dagger})$ are fermionic
and bosonic operators with canonical commutation relations.
The tunneling Hamiltonian finally reads:
\begin{equation}
H_{\mathrm T,\alpha}=\sum_{k_\alpha,n}t\,(c^\dagger_{k_\alpha} c_n+\mathrm{H.c.})\;,\label{tunham}
\end{equation}
where we suppose that the hopping matrix elements $t$ between the two leads and the dot
are of the same order.

We should point out that the coupling to the electronic states of
the SWNT does not significantly affect fullerene vibrations, which
are mostly determined by fullerene {\em confinement} by structural
defects on the carbon nanotube. This is true even though the
coupling to each single electronic state may be strong and depend on
the fullerene position. However, when evaluating the total fullerene
energy shift caused by coupling to a very large number of different
electronic states, the coordinate dependence averages out and such a
coupling does not contribute to the mechanical force.

The Hamiltonian $H_\mathrm{QD}$ can be diagonalized by means of a unitary transformation, see {\em e.g.} \cite{Maha}:
$\overline{H}_\mathrm{QD}=e^S H_\mathrm{QD}\, e^{-S}$, with
$S= \sum_n (V_n(L,l)/\hbar\omega_0)c_n^{\dagger}c_n (b^{\dagger}-b)$. The result
of this transformation is
\begin{equation}
\overline{H}_\mathrm{QD}=\sum_n(\varepsilon_n-\Delta) c_n^{\dagger}c_n +\hbar\omega_0 b^{\dagger}b\;,\label{qdhamtra}
\end{equation}
where $\Delta= \lambda_n^2 \hbar\omega_0$ is the polaronic shift of the resonant energy level and
 $\lambda_n=V_n(L,l)/(\hbar\omega_0)$ is the electron-vibron coupling constant. The tunneling Hamiltonian is
also transformed by the unitary trasnformation and it reads
\begin{equation}
\overline{H}_{\mathrm T,\alpha}=\sum_{k_\alpha,n}t\,(c_{k_\alpha}^{\dagger} c_n\, X+\mathrm{H.c.})\,\label{tunhamtra}
\end{equation}
where the operator $X$ is defined as
\begin{equation}
X=\exp\left [ -\frac{V_n(L,l)}{\hbar\omega_0}(b^\dagger-b)\right]\;.
\end{equation}


\section{Resonant tunneling}\label{transport}

Resonant electron tunneling through a vibrating quantum dot was
considered in many papers (see e.g. Refs.~\cite{9,10} and references
therein). For strong electron-phonon interaction, two
non-perturbative effects determine the low-temperature electron
transport - (i) a polaronic shift of the resonant energy level,
$\varepsilon_n\rightarrow \varepsilon_n - \lambda^2\hbar\omega_0$
and (ii) a "polaronic blockade" (exponential suppression) of the peak
conductance at low temperatures, $G_{\lambda}\propto \exp(-\lambda^2)$.
In our case, the coupling "constant"
$\lambda_n$ is a level-dependent
quantity and it strongly fluctuates from level to level. Such fluctuations are well
visible in FIG. 2.
\begin{figure}
\centering
\includegraphics[scale=0.7]{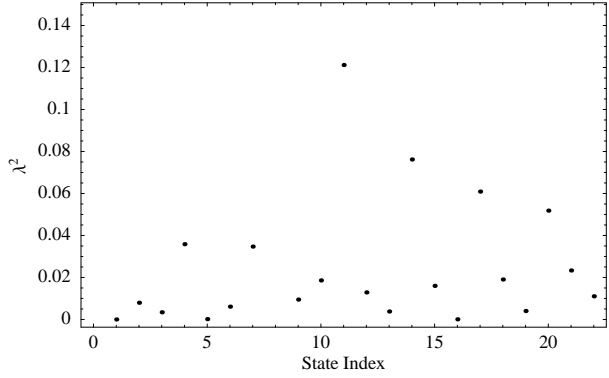}
\caption{Behaviour of the polaronic coupling constant $\lambda_n^2$ as
a function of state index $n$. The coupling constant strongly fluctuates
since it is proportional to the derivative of the energy spectrum.
The corresponding value for the 8-th level, of the order of unity,
is off the scale. The Fermi level is between the 10-th and the 11-th
levels. We stress that we do not intend to show the actual and
absolute numerical values of the couplings $\lambda$. These
"constants" can be obtained by fitting the experimental data to
theory, as in \cite{utko2}. FIG. 2 only aims to illustrate the
strong level-to-level fluctuations of the coupling constant. In the
case of $N$ independent molecules in the peapod, the average value
of the coupling constant can be $N$ times larger than the values
shown in the plot. In particular, the values of the coupling
constant $\lambda$ shown in the plot have been evaluated for a
sample whose length is ten nanometres, and with a LUMO energy and a
hybridization energy of the same order as the Fermi energy. These
are the same values used to evaluate the spectrum in FIG. 1.}
\end{figure}
We observe that these coupling constants being proportional to the derivative
of the energy spectrum exhibit much more pronounced fluctuations then
the energies themselves. The mean value of $\lambda_n^2$ for strong
backscattering potential ($U(k_F)\sim \lambda_\mathrm F^{-1}$) can be
estimated as
$\bar{\lambda}_n^2\sim(\varepsilon_\mathrm Fx_0/\hbar\omega_0L)^2$. So,
for the case of $N$ independent scatterers $N=L/d_\mathrm c,(\;d_\mathrm c$  is the
characteristic range of the longitudinal confinement potential) the total
polaronic shift will scale as $1/L$. Since the charging energy of
our 1D system also scales as $1/L$, it will be difficult to single
out polaronic effects in the distribution of peak spacings
as a function of gate voltage. Besides, the irregular character of
level spacings could be caused by different physical effects, {\em e.g.},
by electron scattering due to imperfections (holes) of the nanotube surface.
Is there another manifestation of low-energy electro-mechanical
effects in resonant electron tunneling?

\subsection{Electric current}

In order to study the peak to peak fluctuations in the electrical conductance, we first
recover in this section the electric current flowing through the system.

We start from the well known Meir-Wingreen formula, \cite{Meir, Meir2},
\begin{equation}
I(V)=-\frac{2 e}{h}\int d\varepsilon \,[f_\mathrm{L}(\varepsilon)-f_\mathrm{R}(\varepsilon)]\Im \mathrm m\{\mathrm{Tr}[\Gamma
G_\mathrm{QD}^\mathrm{ret}(\varepsilon)]\}\;,\label{curr}
\end{equation}
where $f_\mathrm{L/R}(\varepsilon)=[\exp(\varepsilon-\varepsilon_\mathrm{F}\mp eV/2)+1]^{-1}$ are the Fermi functions for the left and right reservoirs, $G_\mathrm{QD}^\mathrm{ret}(\varepsilon)$ is the retarded Green function of the dot and
$\Gamma=\Gamma_\mathrm{L}\Gamma_\mathrm{R}/(\Gamma_\mathrm{L}+\Gamma_\mathrm{R})$.
$\Gamma_\mathrm{L/R}=2\pi t^2 \rho_\mathrm{L/R}$ are the widths of the resonance due to tunneling from the left
or right lead, and $\rho_\mathrm{L/R}$ is the density of the states (DoS) in the reservoirs.

In order to evaluate the current in Eq. (\ref{curr}), one needs the
Green function of the dot which takes into account, at least in
principle, not only all the possible processes in the dot, as
inelastic processes and multiple scattering processes, but also the
effects of the vibrons on the reservoirs, \cite{10,Meir}. To get
analytical results, we will treat the leads, as in \cite{10}, as
unaffected by the vibrons modes in the dot. Besides, we suppose that
the bandwidth in the contacts is much larger than both the resonance
width and the vibron energy quantum, that is we will evaluate the
electron part of Green function in the so called wide-band limit,
\cite{Wing}.

Under such assumptions, the retarded single particle Green function can easily be evaluated, see also \cite{Maha}, and it reads
\begin{eqnarray}\label{Grfun}
G_\mathrm{QD}^\mathrm{ret}(t)
&\simeq& -i \Theta(t)\exp\left [-\frac{i}{\hbar}\left (\varepsilon_n-\Delta-i\frac{\Gamma_\mathrm{L}+\Gamma_\mathrm{R}}{2}
\right )t\right]\exp\left [-\lambda^2(2 n(\beta\hbar\omega_0)+1)\right]\nonumber\\
&\times&\sum_{l=-\infty}^\infty I_l\{2\lambda^2[n(\beta\hbar\omega_0)(n(\beta\hbar\omega_0)+1)]^{1/2}\}\exp\left [l\hbar\omega_0\left(i \frac{t}{\hbar}-\frac{\beta}{2}\right)\right],
\end{eqnarray}
where $n(x)=(e^{x}-1)^{-1}$ is the Bose distribution, $\beta=1/k_\mathrm B T$, $I_l$ is the modified Bessel function
of the first kind, and for sake of simplicity, we have omitted the index $n$ from the coupling constant $\lambda$.
Equation (\ref{Grfun}) can be easily Fourier transformed and the expression of the current written as in \cite{10}
\begin{eqnarray}
I(V)&=&\frac{e}{h}\int d\varepsilon [f_\mathrm{L}(\varepsilon)-f_\mathrm{R}(\varepsilon)]
e^{-\lambda^2(2 n(\beta \hbar \omega_0 )+1)}\nonumber\\
&\times&\sum_{l=-\infty}^\infty I_l\{2\lambda^2[n(\beta \hbar \omega_0 )(n(\beta \hbar \omega_0)+1)]^{1/2}\} e^{-l \hbar\omega_0\beta/2}\,
{\cal T}_\mathrm{BW}(\varepsilon,l)\;,\label{curr1}
\end{eqnarray}
where
\begin{equation}\label{lorent}
{\cal T}_\mathrm{BW}(\varepsilon,l)=\frac{\Gamma_\mathrm{L}\Gamma_\mathrm{R}}{(\varepsilon+l\hbar\omega_0
+\Delta_\mathrm{pol})^2+
\left (\frac{\Gamma_\mathrm{L}+\Gamma_\mathrm{R}}{2}\right)^2}\;,
\end{equation}
with $\Delta_\mathrm{pol}=\Delta-\varepsilon_n$.

\subsection{Electric conductance}

A simple expression of the conductance can be derived from Eq. (\ref{curr1}) in the linear regime, $eV\rightarrow 0$, and weak coupling to the leads, $\Gamma_\mathrm{L}+\Gamma_\mathrm{R}\ll \mathrm{min}\{T, \hbar\omega_0\}$. In addition to the latter condition,
in order to have polaronic states, it is necessary, physically, that $\Gamma\ll T\ll \lambda^2 \hbar \omega_0$, as well.
These conditions state that the time spent by the electron in the dot must be
larger than the characteristic time needed to form a polaron. Under these assumptions, the Lorentzian in Eq. (\ref{lorent})
is just a $\delta$-function centered at $\varepsilon=\varepsilon_n-\Delta-l\hbar\omega_0$.
Then, the integral in Eq. (\ref{curr1}) can be easily evaluated and the current reads
\begin{eqnarray}
I(V)&=&2\pi e\frac{\Gamma}{h}[f_\mathrm{L}(\varepsilon_n-\Delta-l\hbar\omega_0)-f_\mathrm{R}(\varepsilon_n-\Delta-l\hbar\omega_0)]\nonumber\\
&\times&e^{-\lambda^2(2 n(\beta \hbar \omega_0 )+1)}\sum_{l=-\infty}^\infty I_l\{2\lambda^2[n(\beta \hbar \omega_0)(n(\beta \hbar \omega_0)+1)]^{1/2}\} e^{-l \hbar\omega_0\beta/2}\;.\label{curr2}
\end{eqnarray}
Since we are interested in the conductance peak height, we consider the physical condition where, by
means of the gate voltage, the levels in the dot are tuned such that $\varepsilon_\mathrm F=\varepsilon_n-\Delta$.
Then, in the limit $eV\rightarrow 0$, a simple analytical expression can be written:
\begin{equation} \label{GT2}
G_{\lambda}(T)=G_{\lambda=0}(T)F_{\lambda}(\hbar\omega_0/k_\mathrm BT)  ,
\end{equation}
where $G_{\lambda=0}(T)\simeq (\pi/2) G_0\Gamma/k_\mathrm BT$ is the standard
resonance conductance of a single-level quantum dot ($G_0=e^2/h$ is the
conductance quantum) and the function
$F_{\lambda}(x)$, with $x=\hbar\omega_0/k_\mathrm BT$, is represented as a series
\begin{eqnarray} \label{Fx}
F_{\lambda}(x)&=&\exp\{-\lambda^2[1+2n(x)]\}
\sum_{l=-\infty}^{\infty}\frac{\exp(-lx/2)I_l[2\lambda^2\sqrt{n(x)(1+n(x))}]}
{\cosh^2(lx/2)} .
\end{eqnarray}
It is easy to check that $F_{\lambda=0}(x)=1$.

For strong electron-vibron coupling ($\lambda\gtrsim 1$), polaronic
effects can significantly suppress the conductance at low temperatures, ($\Gamma\ll
k_\mathrm BT\ll\hbar\omega_0$), for which one finds $F(x)\simeq\exp(-\lambda^2)$. In this regime,
the usual $\sim 1/T$ behaviour is recovered for the conductance, but the latter is strongly
suppressed because of the factor $\exp(-\lambda^2)$. This is the signal of the
reduced probability for an electron to tunnel from a bare electronic
state in the reservoir to a polaronic state in the dot. As soon as the
temperature becomes of the order of $\hbar \omega_0$, thermally excited vibronic modes of the
fullerene molecule appear. The characteristic energy of fullerene thermal vibrations is
$n(\beta \hbar\omega_0)\hbar\omega_0$, $n(x)$ being the Bose distribution. This energy should be compared
with polaronic energy shift $\lambda^2\hbar\omega_0$ induced by fullerene displacement.
The corresponding ratio determines the relative strength of destructive influence
of thermal vibrations on polaronic effects. Indeed, at temperatures
$\hbar\omega_0\ll k_\mathrm BT\lesssim\lambda^2\hbar\omega_0$, by keeping the first l-independent term in the
 asymptotic expansion of the Bessel function $I_l$,
$F(x)\simeq((\sqrt{\pi/ x}\,)/\lambda)\exp(-\lambda^2x/4)$, and the conductance reads,
%

%
\begin{equation} \label{GT1}
G_{\lambda}(T)\simeq
G_0\frac{\Gamma}{\lambda\hbar\omega_0}\left(\frac{k_\mathrm BT}{\hbar\omega_0}
\right)^{-1/2}\exp\left(-\lambda^2\frac{\hbar\omega_0}{4k_\mathrm BT}\right) .
\end{equation}
\begin{figure}
\centering
\includegraphics[scale=0.45]{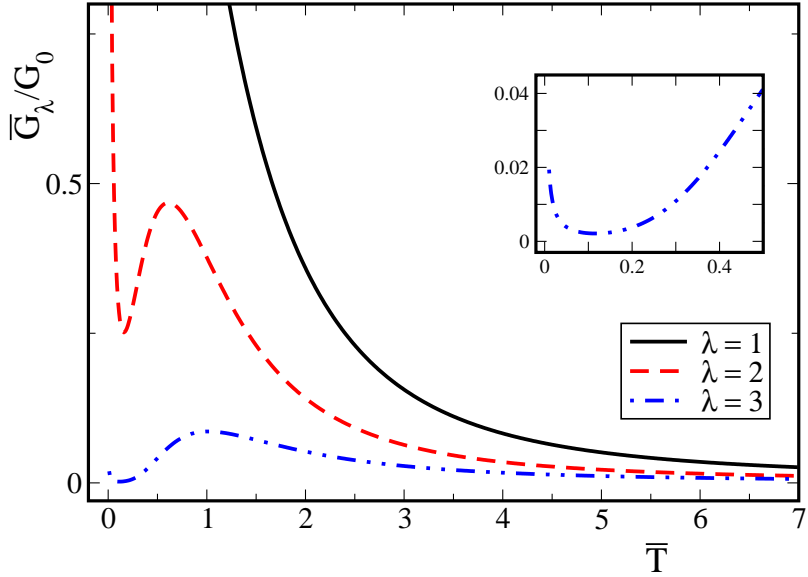}
\caption{ (Color online) Behavior of the conductance ${\overline G}_\lambda (T)=G_\lambda \hbar\omega_0/\Gamma$ as a function of the renormalized temperature ${\overline T}=k_\mathrm B T/\hbar\omega_0$ and for three different values of the
coupling constant $\lambda$. An anomalous temperature dependence of the resonance peak height is well visible in the region $k_\mathrm B T\lesssim \lambda^2\hbar\omega_0$.
Inset: the behaviour of the conductance
for strong coupling, $\lambda=3$, is shown in a narrow region close to zero temperature. The behaviour is dominated by
$\exp (-\lambda^2)/T$ close to zero. Then, the dominant term is proportional to $\sim (1/\sqrt{T})\exp (-\lambda^2\hbar\omega_0/4k_\mathrm B T)$. }
\end{figure}
The latter equation indicates that for temperatures larger than $\hbar\omega_0$, a
competition arises between the vibrons and the polarons, giving rise to a non-monotonic
behaviour.
At temperatures
$k_\mathrm BT\gg\lambda^2\hbar\omega_0$, when the polaronic blockade is completely
lifted, the conductance scales as $1/T$, as can be numerically verified.
At such high temperatures, the
polaronic effects get completely destroyed by the thermally excited
vibrons. The temperature lifting
of the polaronic blockade results in a {\em non-monotonic} temperature dependence
of the conductance peak. This behavior is shown in FIG. 3, where we
plot the temperature dependence of the amplitude of the Coulomb blockade conductance peak
for different values of the electron-vibron coupling $\lambda^2$.
It is clearly seen from the figure that polaronic effects are
pronounced only for strong electron-vibron interaction ($\lambda\gtrsim 1$),
and in the temperature region $k_\mathrm BT\lesssim \lambda^2\hbar\omega_0$
anomalous temperature dependence of resonance peaks appears.
The observation in an experiment of a low-temperature scaling
which differs from the usual $1/T$-dependence for narrow Coulomb blockade conductance peaks,
in general, indicates
that the level width is not the only relevant energy scale in the problem.
An anomalous temperature behavior and $1/\sqrt{T}$-scaling is
the signature of strong polaronic effects, as experimentally observed, \cite{utko2}.

\section{Polaronic effects in long-range scattering limit}\label{Graphene}

The polaronic effects discussed in the previous Sections appear due
to strong backscattering of the SWNT electrons by the
fullerene-induced potential, which was assumed to be short-range.
Here we want to consider the opposite limit of a long-range
potential, where we can also obtain an analytical solution. In this
case the real band structure of the SWNT electrons must be
considered. It is known that the specific band structure (Dirac-like
spectrum) leads to a new type of backscattering by even a long-range
potential. In this Section we will study the influence on polaronic
effects of this type of backscattering, which is closely related to
the chiral properties of the carbon nanotube (the chiral angle).


%

%

The band structure of graphene in the vicinity of the Fermi energy
(these points in the Brillouin zone of 2D graphite are usually
labelled K and K') is described by the Dirac Hamiltonian for massless
fermions $ H_\mathrm D=\hbar
v_\mathrm F\vec{\sigma}\vec{p}$, \cite{13} (see Refs.~\cite{exp} where a Dirac-like energy
spectrum in graphene was confirmed in experiments). Here, $\vec{p}$ is the 2D electron momentum,
$\vec{\sigma}$ are the Pauli 2$\times$2 matrices which describe the
chiral properties (pseudospin) of particles and holes in each
conduction band. The chiralities $\vec{\sigma}\vec{p}/|\vec{p}|=\pm
1$ of particles and holes are opposite in the same band and
particles (holes) in different bands are characterized also by opposite
chiralities. Conservation of chirality in scattering by a scalar
potential leads to various unusual phenomena in graphene that are currently
discussed in the literature (see, {\em e.g.}, Ref.~\cite{14}). The wrapping
of a graphene sheet into a SWNT results in a 1D relativistic form of the
electron dispersion, which depending on the chiral indeces of the
nanotube, describes either massless (metallic SWNT) or massive
(semiconducting SWNT) Dirac fermions $\varepsilon(q)=\pm\hbar
v_\mathrm F\sqrt{q^2+\Delta_m^2}$, where $\Delta_m\propto 1/R_\mathrm{NT}$ is the
gap in electron spectrum, and $R_\mathrm{NT}$ is the radius of the nanotube.

Now, we will treat the hybridization-induced scattering potential as
a long-range potential \cite{4} and reconsider the problem of
resonant electron tunneling using the Dirac-like electron spectrum.
Scattering of electrons by a long-range potential ($\delta k\ll
k_\mathrm F$) leaves the quasiparticles in the same band and thus does not
mix particles (or holes) with opposite chiralities. In this case
the chiral properties of electrons in a SWNT play a significant role and
should be taken into account.  Using the same approach as in the
Ref.~\cite{4}, we consider the spectrum problem for a short
peapod and discuss possible polaronic effects in resonant electron
transport through this system.

At the boundaries of a nanotube, $x=0,L$, the
scattering potential is sharp and this results in strong
backscattering $\delta k\simeq 2k_F$ by the boundaries. The spinor
wave function of an electron in a nanotube now takes the form $
\Psi(x)=e^{ik_\mathrm Fx}\Psi_{+}(x) + e^{-ik_\mathrm Fx}\Psi_{-}(x)$ , where
subindices (+,-) label two identical bands of the energy spectrum. The
general boundary condition for a finite-sized system of Dirac
fermions is the absence of current through the boundaries. This
condition is satisfied in our case if we assume perfect
reflection at the boundaries, $L^{\pm}\leftrightarrow R^{\mp}$,
where $R^{\pm}(L^{\pm})$ denote right(left)-moving particles in
the "+" or "-" band.

In reference \cite{4}, the scattering matrix for a long-range
potential was calculated using the Dirac equation. Now, we would
like to incorporate this result in order to consider resonant
tunneling through the one dimensional wire. This can be easily done
in the limit when the distances from the scatterer to the ends of
the wire are much bigger than the
spatial extension of the potential. Hence, if one is not interested
in the exact position of the resonant level, then the resonant
transmission through the given level can be obtained in the
approximation where the scatterer can be treated as a point-like
defect. Such a defect will be characterized by the same scattering
matrix obtained for the electrons propagating in an infinite wire as
in \cite{4}.

The local ($\delta$-function) potential at $x=l$  now represents a
long-range imperfection which scatters fermions only within its
own band. The spectral problem can be solved by the standard
method of finding the solution of the Dirac equation in the ranges
$x\neq l$ and matching them at $x=l$ by using the properties of
the Dirac equation in a $\delta$-function potential. In the general case of
massive Dirac particles, the spectrum equation is cumbersome and
lengthy. We are interested in transport through a metallic SWNT,
that is in the case of a massless Dirac spectrum, $\Delta_m=0$. In
this limit, $\varepsilon=\pm\hbar v_\mathrm Fq$, the spectrum equation
can be represented in the simple form
\begin{equation} \label{Ds}
t(\theta,\varphi)\cos(2k_\mathrm FL) + r(\theta,\varphi)\cos[2q(L-2l)] -
\cos(2qL-\varphi)=0 ,
\end{equation}
where $t(\theta,\varphi), r(\theta,\varphi)$ are the transmission
and reflection coefficients for the $\delta$-function potential scattering problem by massless
chiral particles. In this case,
\begin{equation} \label{rc}
r(\theta,\varphi)=\sin^2\theta\sin^2\frac{\varphi}{2}\;,
\end{equation}
where $\theta$ is the chiral angle of the nanotube and
\begin{equation} \label{phi}
\tan\varphi = \frac{2\hbar v_\mathrm FV(\varepsilon)}{(\hbar
v_\mathrm F)^2-V^2(\varepsilon)}\;\;\;,
\;\;\;V(\varepsilon)=V_\mathrm h(\varepsilon)\;.
\end{equation}
We see from Eqs. (\ref{Ds})-(\ref{phi}) that for a massless Dirac
spectrum the reflection coefficient vanishes for $\theta=0$, that is
for armchair nanotubes. The corresponding spectrum is uniform
with level spacing $\pi\hbar v_\mathrm F/2L$, and the total energy shift,
$\hbar v_\mathrm F\varphi/2L$, is induced by forward electron scattering by the
long-range potential. The absence of electron backscattering in
non-chiral SWNTs (formally due to conservation of chirality) has
repeatedly been discussed in the literature (see e.g. Ref.~\cite{14}). In
general, the suppression of backscattering in metallic SWNTs as
compared to semiconducting nanotubes (where the presence of a gap in
the spectrum mixes states with opposite chiralities) is often used
to explain the remarkably good conducting properties of
long metallic carbon nanotubes, \cite{15}.

For the spectrum in Eq.~(\ref{Ds}), one can easily estimate the
electron-vibron coupling as $\lambda\sim
r(\theta,\varphi)v_\mathrm Fx_0/\omega_0L^2$. Polaronic effects are
determined by the square of this parameter, which is smaller by a factor
$(\lambda_\mathrm F/L)^2$ then the analogous coupling expected for a
short-range hybridization potential.

The origin of such a weak polaronic
coupling constant resides in the strongly reduced backscattering processes, since
polaronic effects are extremely sensible to that. Then, in
transport measurements, where electronic backscattering with a large momentum transfer is
expected because of the same order of magnitude of the electron Fermi
wavelength and the fullerene diameter, the description of
the hybridization-induced scattering potential as a long-range potential
is not adequate. In this case, short-range approximation
(sections \ref{spectrum} and \ref{transport}) is more appropriate for description of
polaronic effects.

\section{Conclusions}

In conclusion, we have showed that the elastically soft subcomponent of peapods
(fullerene molecules trapped inside a SWNT) can strongly influence
low-temperature properties of resonant electron transport through
short metallic peapods. If encapsulated $C_{60}$'s do modify the
electronic structure of metallic nanotubes, (the effect was already
observed for semiconducting peapods \cite{3}), the mechanical degree
of freedom associated with fullerene dynamics will influence, via
polaronic effects, both the distribution of conductance peak amplitudes
and peak spacings as a function of gate voltage. In our
model we considered independent vibrations of individual fullerenes
(a disordered chain of $C_{60}$'s inside a nanotube). This model
predicts strong fluctuations and anomalous low-temperature behavior of Coulomb
blockade conductance peaks. Such fluctuations have been experimentally observed, \cite{utko2}.

One more question remains open, since our Hamiltonian, Eq.
(\ref{hamilt}), does not include the effect of the scattering
potentials due to the holes on the external shell on the nanotube
electrons. In principle, the scattering of the electrons by these
holes should be included. Our approximation, which neglects these
effects, is based on two equally important observations.

The first observation is that there is experimental evidence, \cite{Utko}, to
suggest that electron transport through a peapod in a two-terminal conduction geometry,
and in presence of a gate electrode, is ballistic over the entire sample (whose length
is around $\sim 400$ nm). In other words, the holes on the nanotube shell do not hinder
ballistic electron transport.

The second observation is the following: the inclusion in the model of any
topological defects (holes) in the tube shell
would certainly affect the spectrum of the electrons in the nanotube and the level spacing.
However this would only result in a renormalization of the coupling constant $\lambda$, which in our model
is a fitting parameter. The anomalous temperature dependence of conductance induced by vibrational
dynamics of encapsulated fullerenes will not be affected.

Finally, we have also showed that the description of the hybridization-induced scattering
potential as a long-range potential is not adequate to describe polaronic effect.
%
%

\section{Acknowledgments}

We thank L. Y. Gorelik for fruitful discussions. This work was
supported by the Royal Swedish Academy of Sciences (KVA), by the
Swedish Research Council (VR) and in part by EC FP6 funding
(contract no. FP6-2004-IST-003673, CANEL). IVK gratefully
acknowledges financial support from the joint grant of the
Ministries of Education and Science in Israel and Ukraine and from
the  grant "Effects of electronic, magnetic and elastic properties
in strongly inhomogeneous nanostructures" (National Academy of
Sciences of Ukraine), and the hospitality of the Department of
Physics at the University of Gothenburg and the Department of
Physics and Astronomy at Tel Aviv University.

\end{document}